\documentclass[aps, pra, reprint, floatfix, showpacs, nofootinbib]{revtex4-1}
\usepackage{graphicx}
\usepackage{amsmath,amssymb}
\usepackage{color}
\newcommand{\doublet}{$^2\Sigma^+$}
\newcommand{\quartet}{$^4\Sigma^+$}

\begin{document}

\author{Dariusz K\c{e}dziera}
\email{teodar@chem.umk.pl}
\affiliation{Department of Chemistry, Nicolaus Copernicus University, 7 Gagarin Street, 87-100 Toru\'n, Poland}

\author{Piotr S.~\.{Z}uchowski}
\email{pzuch@fizyka.umk.pl}
\affiliation{Institute of Physics, Faculty of Physics, Astronomy and Informatics, Nicolaus Copernicus University, ul.\ Grudziadzka 5/7,       87-100 Toru\'n, Poland}

\author{{\L}ukasz Mentel}
\affiliation{Department of Chemistry, Nicolaus Copernicus University, 7 Gagarin Street, 87-100 Toru\'n, Poland}

\author{Steven Knoop}
\email{s.knoop@vu.nl}
\affiliation{LaserLaB, Department of Physics and Astronomy, VU University, De Boelelaan 1081, 1081 HV Amsterdam, The Netherlands}

\title{{\em Ab initio} interaction potentials and scattering lengths for ultracold mixtures of metastable helium and alkali-metal atoms}

\date{\today}

\begin{abstract}
We have obtained accurate {\em ab initio} \quartet~quartet potentials for the diatomic metastable triplet helium + alkali-metal (Li, Na, K, Rb) systems, using all-electron restricted open-shell coupled cluster singles and doubles with noniterative triples corrections [CCSD(T)] calculations and accurate calculations of the long-range $C_6$ coefficients. These potentials provide accurate {\em ab initio} quartet scattering lengths, which for these many-electron systems is possible, because of the small reduced masses and shallow potentials that results in a small amount of bound states. Our results are relevant for ultracold metastable triplet helium + alkali-metal mixture experiments. 
\end{abstract}


\maketitle

\section{Introduction}

Interactions and collisions involving helium in the metastable triplet 2~$^3$S$_1$ state (denoted as He$^*$) have been regarded for many years as one of the most fascinating in atomic and molecular physics. The enormous amount of internal energy (19.8 eV) allows for Penning ionization, 
which has been extensively exploited in crossed-beam studies \cite{siska1993mbs}. More recently, molecular-beam techniques involving He$^*$ have achieved sufficient resolution in kinetic energy \cite{henson2012oor,lavertofir2014oot,jankunas2014oos} to observe for shape resonances in single partial waves for sub-Kelvin collisions, which are a sensitive probe of the interaction potential. 

Elastic and inelastic collisions at sub-milliKelvin energies are relevant for ultracold trapped He$^*$ gases \cite{vassen2012cat}. Penning ionization greatly limits the lifetime of the trapped gas. However, for a spin-polarized gas of He$^*$ Penning ionization is strongly suppressed due to spin conservation \cite{shlyapnikov1994dka}. Bose-Einstein condensates of $^4$He$^*$ have been realized  \cite{robert2001bec,pereira2001bec,tychkov2006mtb,dall2007bec,keller2014bec}, benefiting from a sufficiently large scattering length of 142~$a_0$ \cite{przybytek2005bft} that allows for efficient evaporative cooling. Degenerate Fermi gases of $^3$He$^*$ are obtained by sympathetic cooling with $^4$He$^*$ \cite{mcnamara2006dgb}, which is efficient due to a very large interspecies scattering length of 496~$a_0$ \cite{przybytek2005bft}. 

One of the unique features of He$^*$ is its simple electronic structure. Thus, it is being explored extensively in atomic physics to confront the state-of-the-art, ultraprecise spectroscopy and the most advanced fundamental theories. For example, precise measurement of transition frequencies and the lifetimes are used to test the quantum electrodynamic (QED) calculations \cite{hodgman2009mha,rooij2011fmi,notermans2014hps}. Similarly, the two-body interaction potential for spin-polarized He$^*$ atoms is so far the only system in ultracold physics for which it is possible to calculate the aforementioned scattering lengths \cite{przybytek2005bft} with an accuracy surpassing the experimental value \cite{moal2006ado}. This contrasts with other systems, such as e.\,g.\, the interaction potential for alkali-metal atoms, for which predicting the scattering length with such accuracy is impossible, and without experimental data, it is essentially unknown.

Recently we have challenged this situation for the interaction between He$^*$ and Rb, and demonstrated a very good agreement between theoretical predictions based on all-electron restricted open-shell coupled cluster singles and doubles with noniterative triples corrections [CCSD(T)] calculations and the scattering length derived from thermalization measurements of an ultracold mixture of $^4$He$^*$ and $^{87}$Rb \cite{knoop2014umo}. The fact that even for so many electrons it is possible for {\em ab initio} calculations to quantitatively predict scattering lengths results not only from the simple electronic structure of He$^*$, but mostly comes from the small reduced mass and shallow interaction potential that give a small number of bound states, making the scattering length less sensitive to uncertainty in the calculated potential (see Ref.~\cite{borkowski2013sli} for the opposite case).  

Motivated by our previous work on He$^*$+Rb \cite{knoop2014umo}, we consider the relevant interaction potentials for He$^*$ and the other alkali-metal atoms. Similarly as in the case of homonuclear He$^*$ collisions, Penning ionization:
\begin{equation}
{\rm He}^*+{\rm A} \rightarrow {\rm He}+{\rm A}^+ +{\rm e}^-\label{PI},
\end{equation}
is suppressed by spin-polarizing He$^*$ and alkali-metal atom A both in the spin-stretched states, such that the total spin and its projection is maximum \cite{byron2010sop}. While the $s$ character of the valence electron of He$^*$ and alkali-metal atoms are alike, we expect the amount of suppression to be similar to that in He$^*$+He$^*$ (in contrast to the other metastable noble gas systems \cite{vassen2012cat}). Here we have assumed that the product ion A$^+$ has zero spin, which is true for the ground state. However, if excited, non-zero spin, (A$^+$)$^*$ states are energetically available, Penning ionization is not spin forbidden, even for the doubly spin-stretched state combination. Among the alkali-metal atoms the (A$^+$)$^*$ states cannot be reached, except for Cs, for which the excitation energy from the neutral to the first excited ionic state is 17.2~eV \cite{sansonetti2009wtp}. Therefore we will discard Cs in this work as a stable ultracold mixture with He$^*$ seems experimentally not feasible and computationally the presence of the energetically available excited ionic channel will complicate the calculations. Note that the above mentioned criteria to suppress Penning ionization exclude most other atomic species as well (except He$^*$+H).

The collision properties for He$^*$ + alkali-metal atom are determined by a \doublet~doublet and a \quartet~quartet potential, however, for the doubly-spin stretched state combination scattering only occurs in quartet potential. Therefore for realizing a stable mixture the properties of the quartet potential are the most relevant, in particular the quartet scattering length. This scattering length determines interspecies thermalization rates and therefore whether sympathetic cooling is efficient or not. Also, for quantum degenerate mixtures it determines, together with the intraspecies scattering lengths, whether the mixture is miscible or immiscible in case of Bose-Bose mixtures \cite{esry1997hft,law1997ssi}, or whether a Fermi core or shell is formed in case of Bose-Fermi mixtures \cite{molmer1998bca}. Similar to the triplet potential in alkali-metal + alkali-metal interactions, the quartet potential is quite shallow, which in combination with the small reduced mass leads to a small number of bound states in the range of 11 to 15. Note that previous experimental work on these kind of collisions systems was based on measuring electron emission spectra in crossed beam experiments \cite{gray1985uop,ruf1987tio,merz1990eat}, which inherently is only sensitive to the doublet potential. Therefore ultracold mixture experiments are the first to explore these quartet potentials.

In this paper we present {\em ab initio} calculations of the quartet potentials of He$^*$ + alkali-metal (Li, Na, K and Rb) systems, where the results for He$^*$+Rb are taken from Ref.~\cite{knoop2014umo}. In Sec.~\ref{theory} we describe the methodology of the calculations, while in Sec.~\ref{potentials} we present the obtained potentials and give a detailed discussion on the accuracy of those potentials. In Sec.~\ref{scatt} we provide the corresponding scattering lengths for all the isotopologues and discuss possible implications for future experiments. Finally, in Sec.~\ref{concl} we conclude and give an outlook. 

Throughout this paper we use Bohr radius, $a_0=5.2917721\times10^{-11}$~m, as a length unit and cm$^{-1}=1.9864455\times 10^{-23}$~J as an  energy unit.
 
\section{Theory}\label{theory}

\subsection{General considerations}

The $s$-wave scattering length $a$ is obtained by solving the 1D radial Schr\"{o}dinger equation with zero angular momentum and vanishing
kinetic energy $E$:
\begin{equation}\label{Schrodinger}
\psi''(r)+\frac{2\mu}{\hbar^2}\left[E-V(r)\right]\psi(r)=0,
\end{equation}
where $\mu$ is the reduced mass, $r$ the internuclear distance and $V(r)$ is the interaction potential. Current quantum chemistry {\em ab initio} methods are able to determine the short-range part of $V(r)$ with an accuracy on the order of few percent, which translates into few cm$^{-1}$ for dispersion-bound systems, like spin-stretched alkali-metal dimers or He$^*$+alkali-metal atom systems. The long-range part of $V(r)$ for alkali-dimers and systems with helium atom can be described very accurately through the van der Waals expansion: the corresponding coefficients for such systems can be obtained with sub-percent accuracy. 

With an appropriate analytical form of $V(r)$ it is possible to explore separately the influence of short- and long-range potential modifications on the scattering length. A good choice of such a function is the so-called Morse/Long-Range (MLR) potential \cite{leroy2006aaa}, which has the form:
\begin{equation}
V(r)=D_e\left(1- \frac{u_{\rm LR}(r)}{u_{\rm LR}(r_e)} \exp\left[-\phi(r) y_p(r)\right]\right)^2-D_e,
\end{equation}
where 
\begin{eqnarray}
u_{\rm LR}(r)&=&\frac{C_6}{r^{6}}+\frac{C_8}{r^{8}}+\frac{C_{10}}{r^{10}}, \\
y_k(r)&=&\frac{r^k-r_e^k}{r^k+r_e^k},\\
\phi(r)&=&\left[1-y_p(r)\right] \sum_{j=0}^4 \phi_j \left[y_q(r) \right]^j + y_p(r) \phi_{\infty}. 
\end{eqnarray}

The free parameters of the MLR potential, determined by fitting, are the $\phi_j$ ($j=0,\dots,4$) coefficients, while the potential well depth $D_e$, equilibrium distance $r_e$ and $\phi_\infty=\log[2 D_e/u_{\rm LR}(r_e)]$ are directly obtained from the {\em ab initio} calculations of the short-range potential. For the long-range part of the potential, $V(r)\stackrel{r\rightarrow\infty}{\rightarrow}-u_{\rm LR}(r)=-C_6 r^{-6}-C_8 r^{-8}-C_{10} r^{-10}$. Note that the statistical error introduced by the analytical fit is much smaller than the systematic uncertainty in the ab initio calculations.

\subsection{Short-range potential}\label{short-range}

For the short-range potential we have used the coupled cluster method~\cite{cizek1966otc} in which the correlated electronic wavefunction is represented by the exponential operator $\exp(T)$ acting on the Slater determinant, where $T=T_1+T_2+\ldots$ is the so-called cluster operator which includes single-, double- and higher-order excitations. A gold standard for weakly bound systems has become the approximate coupled cluster method denoted as CCSD(T), in which one fully includes single- and double-excitations and treats triple-excitations approximately. In this paper we use the open-shell version of CCSD(T) introduced by Knowles {\em et al.} \cite{knowles1993cct}, implemented in the {\sc molpro} program \cite{MOLPRO_brief}.
 
The accuracy of CCSD(T) for predicting binding energies for weakly bound systems is typically on the order of 2-3\% percent~\cite{smith2014bsc}. Here we confirm this accuracy by calculating the $D_e$ parameter using an even more sophisticated method, namely coupled cluster with full triple excitations, CCSDT. In addition, we have incorporated scalar relativistic effects using the Douglas-Kroll-Hess approximation up to the fifth order in external potential (DKH5)~\cite{reiher2004edo}. All electrons have been correlated in these calculations. Interaction energies were calculated using the counterpoise correction scheme of Boys and Bernardi~\cite{boys1970tco}, in which the total monomer energies  calculated in dimer basis sets are substracted from  total energy of dimer.

The accuracy of employed quantum chemistry methodology is also determined by the choice of appropriate gaussian basis set describing the orbitals. Due to computational limitations, the basis set commonly used in quantum chemical calculations are truncated. However, the basis sets that are used in this paper are tailored to systematically improve the correlation energy and allow approximate extrapolation to the complete basis set limit (CBS). By increasing the maximum angular momentum in the basis sets the errors in correlation energy are expected to decrease. Hence, in order to converge the quantum chemical calculations one should obtain the results for a series of basis sets with increasing maximum angular momentum. 

While for most atoms corresponding to the first- and second rows in periodic table a variety of various families of well-optimized, systematically convergent basis sets are available, for heavy alkali-metal atoms (K, Rb, Cs) the choice of basis sets is limited. We have used a sequence of core-valence correlation consistent basis sets developed by Prascher {\em et al.} \cite{prascher2011gbs} for Li and Na, which we will denote as TZ, QZ and 5Z. For K we have used an uncontracted atomic natural orbital relativistic  basis sets \cite{roos2008nra} (denoted as ANO-RCC) with the exponents of $h$ functions taken from $g$ of the same basis (we follow the usual convention to label the gaussian functions with appropriate orbital angular momentum). In each case the basis set have been augmented by a set of two additional diffused functions per shell generated as an even tempered set according to prescription implemented in the {\sc molpro} program. The angular momentum structure of largest basis sets for Li, Na and K is $20s14p10d8f6g1h$, $22s20p10d8f6g2h$, $23s18p7d4g2h$, respectively. 

A proper choice of helium basis set is also essential: the optimal basis sets for the metastable triplet state have entirely different character compared to basis sets for ground state helium \cite{przybytek2005bft}. Hence, we have decided to build a new basis set that takes into account diffused character of He$^*$. To this end we have optimized a new set of exponents according to a following procedure. The starting point was an uncontracted ANO-RCC basis set for ground-state helium, i.\,e.\, $9s4p3d2f$. Exponents were re-optimized to minimize the total energy in the DKH5 relativistic method (at the full configuration interaction level) and extended to $15s8p5d4f2g$, which gives less than 1~$\mu$Hartree convergence. Then the lowest exponents for each shell were again augmented and re-optimized for the total energy of helium dimer quintet state at the equilibrium distance. The final helium basis set has an angular momentum structure of $17s10p7d5f4g$. 

With such basis sets we have performed test calculations for the spin-polarized homonuclear systems He$^*_2$, Li$_2$, Na$_2$ and K$_2$ to assess their performance near the equilibrium distance. We have obtained very good agreement of $D_e$ parameters with experimental results: for He$^*_2$ we have obtained 1042.3~cm$^{-1}$ which is very close to estimated limit of complete basis set for CCSD(T) method (1042.9~cm$^{-1}$)~\cite{przybytek2005bft} and about 5~cm$^{-1}$ shallower than the theoretical potential obtained with a full configuration interaction method. For Li$_2$, Na$_2$ and K$_2$ systems CCSD(T) values calculated with basis sets described above are respectively, 330.5~cm$^{-1}$, 172.3~cm$^{-1}$ and 247.5~cm$^{-1}$, which can be compared with experimental values of 333.69~cm$^{-1}$, 174.96~cm$^{-1}$ and 255.017~cm$^{-1}$~\cite{linton1999thl,ho2000rkt,pashov2008sot}. Clearly CCSD(T) with current basis sets systematically underestimates the well depths, but the deviation from the benchmark values are small. 

\subsection{Long-range potential}\label{long-range}

The $C_6$ van der Waals coefficients of a general A+B system can be obtained by integration of the dipolar dynamic polarizabilities $\alpha$ over the imaginary frequencies \cite{knowles1987asm}:
\begin{equation}\label{C6formula} 
C_6^{\rm AB} = \frac{3}{\pi} \int \alpha^{\rm A} (i\omega ) \alpha^{\rm B} (i\omega)  d \omega.
\end{equation}
For He$^*$ we have used the polarizabilities that are obtained from explicitly correlated gaussian wavefunctions with an accuracy on the order of 0.1\% for the zero-frequency \cite{knoop2014umo,puchalskiprivate}. In case of alkali-metal atoms we have used the dynamic polarizabilities at imaginary frequencies given by Derevianko {\em et al.} \cite{derevianko2010edp}. These polarizabilities give homonuclear $C_6$ coefficients with an accuracy of 0.14\%, 0.25\%  and 0.4\% for the Li+Li, Na+Na and K+K systems, respectively. Hence, the inaccuracy of $C_6$ in heteronuclear systems determined as $\sqrt{\frac{1}{2}(\delta_{\rm A}^2+\delta_{\rm B}^2)}$~\cite{derevianko2010edp} (where $\delta_{\rm A}$ and $\delta_{\rm B}$ are unsigned errors of pertinent static polarizabilities of monomers A and B, respectively)  is always smaller than 0.25\%. Therefore the uncertainty in the long-range part of the potential is much smaller than that of the short-range part obtained from CCSD(T), and one can treat the long-range part of the interaction as fixed by theory. Note that our $C_6$ coefficients agree with the values obtained by Zhang {\em et al.} \cite{zhang2012lri} to better than 0.1\%, which suggests that their estimated uncertainty of 1-5\% is too conservative. 

For the $C_8$ and $C_{10}$ coefficients we have used the values calculated by Zhang {\em et al.} \cite{zhang2012lri} for which the authors estimated an uncertainty of about 1-10\%. However, our study of He$^*$+Rb system has shown that the error bounds for the $C_8$ coefficient given by Zhang {\em et al.} was also far too conservative \cite{knoop2014umo}.

\section{{\em Ab initio} potentials}\label{potentials}

\subsection{Recommended {\em ab initio} potentials}

\begin{figure}
\includegraphics[width=8.5cm]{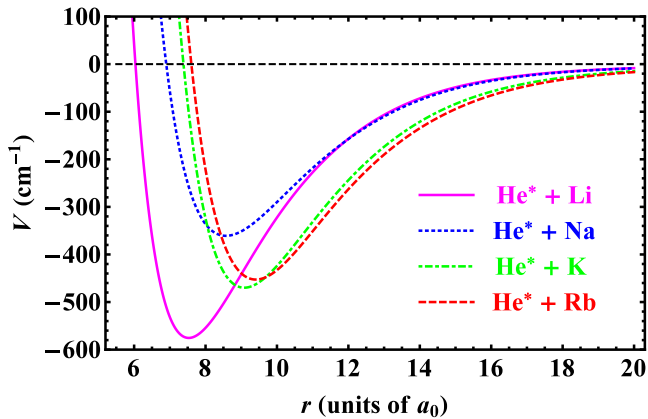}
\caption{(Color online)
\label{quartetpotentials} Results of the \textit{ab initio} calculations of the potential energy curves of the quartet ($^4\Sigma^+$) He$^*$ + alkali-metal systems.}
\end{figure}

The results of our calculations for He$^*$+(Li, Na, K, Rb) quartet potentials are shown in the Fig.~\ref{quartetpotentials}; the values of the potential well depth $D_e$, the equilibrium distance $r_e$ and the $C_6$ coefficient are given in Table~\ref{potentialtable}. A complete list of the parameter values of the MLR potential are given in Appendix~\ref{app} (Table~\ref{MLRvalues}). The uncertainty of the {\em ab initio} calculations is predominantly translated in an uncertainty in $D_e$, while the uncertainty in $r_e$ is smaller than 0.01~$a_0$. 

\begin{table}
\caption{Key parameters of potential energy curves of quartet ($^4\Sigma^+$) He$^*$ + alkali-metal systems, including equilibrium distance $r_e$, the potential depth $D_e$ and the $C_6$ long-range coefficient. The uncertainty of the CCSD(T) calculations is reflected in the error bars on $D_e$ (see text).}\begin{ruledtabular}
\begin{tabular}{l|ccc}
system     	 &  $r_e$~($a_0$)	&	$D_e$~(cm$^{-1}$)		& $C_6$~(cm$^{-1}$$a_0^6$)    \\   \hline
He$^*$ + Li  &  7.53   				& 575$_{-1}^{+4}$   	& $4.5782\times10^8$   \\  
He$^*$ + Na  &  8.57   				& 361$_{-1}^{+4}$   	& $4.7723\times10^8$   \\ 
He$^*$ + K   &  9.08   				& 470$_{-1}^{+9}$ 		& $7.7314\times10^8$   \\  
He$^*$ + Rb  &  9.41   				& 453$_{-1}^{+8}$	 		& $8.4673\times10^8$  \\ 
 \end{tabular}
\end{ruledtabular}
\label{potentialtable}
\end{table}

The overall qualitative pattern of the potentials resembles the triplet states of homonuclear alkali-metal systems: the one containing Li has the deepest well, K and Rb are slightly shallower than Li, and Na is anomaly shallower than the other potentials. This pattern can be explained by subtle interplay between attractive dispersion forces and Pauli repulsion (exchange energy). The Pauli repulsion increases with system size as the atoms are systematically more diffused and have systematically larger radius. Increase in exchange energy is reflected in monotonic increase of equilibrium distance of dimers. On the other hand, the dispersion interaction gives rise to the attraction of atoms in the system and its magnitude correlates with the $C_6$ coefficients. Li and Na have comparable dispersion interactions, but a much larger exchange repulsion in the case of Na results in a much smaller well depth. On the other hand, K and Rb exhibit much stronger dispersion interaction, thus their well depths are noticeably larger than that of Na. 

\subsection{Accuracy of the {\em ab initio} potentials}

To provide tests of the potential accuracy in the minimum region we have performed additional calculations for He$^*$+(Li, Na) using a coupled-clusters approach with full triple excitations \cite{watts1990tcc} (CCSDT) for basis sets with maximum angular momentum limited to $f$ (for Li and Na) and $d$ (for He$^*$) functions. CCSDT calculations were performed using the {\sc cfour} quantum chemistry code \cite{cfour_brief}. The results are shown in the Table~\ref{LiNatable}, and are discussed below. Note that a detailed description of the accuracy of the He$^*$+Rb potential is given in Ref.~\cite{knoop2014umo}.

\begin{table}
\caption{Interaction energies of quartet states of the He$^*$+(Li, Na) systems calculated for the distance corresponding to $r_e$ of recommended potential (7.53~$a_0$ and 8.57~$a_0$, respectively) using CCSD(T) and CCSDT levels of coupled-clusters theory. We report also test calculations with various basis sets and levels of frozen-core approximation. See Sec. \ref{short-range} for notations regarding the basis sets. }\begin{ruledtabular}
\begin{tabular}{l|ccc}
basis set 	 &  level   & active electrons	 & $V$($r_e$)~(cm$^{-1}$)	  \\   \hline
\multicolumn{4}{c}{He$^*$+Li} \\ \hline
TZ 	& CCSDT    	& all	&		572.6   \\ 
TZ  & CCSD(T)  	& all	&   569.2 \\  
TZ  & CCSDT   	& 3 	&  	578.3  \\ 
TZ 	& CCSD(T)  	& 3 	&   576.6  \\  
QZ 	& CCSD(T)  	& all	&  	574.5  \\  
\textbf{5Z} 	& \textbf{CCSD(T)}  	& \textbf{all} &  	\textbf{575.4}  \\  
CBS & CCSD(T) 	& all	&  	576.1	  \\ \hline 
\multicolumn{4}{c}{He$^*$+Na} \\ \hline
TZ 	& CCSDT    	& 7		&   352.0   \\ 
TZ  & CCSD(T)  	& 7  	&   350.2 \\  
QZ 	& CCSD(T)  	& all	&  	362.2  \\  
\textbf{5Z} 	& \textbf{CCSD(T)}  	& \textbf{all}	&  	\textbf{360.8}  \\  
CBS & CCSD(T) 	& all	&  	359.3	  \\  
 \end{tabular}
\end{ruledtabular}
\label{LiNatable}
\end{table}

\subsubsection{Basis set convergence}
\label{basissetconv}
We have studied the basis set convergence for all systems. To this end we studied the dependence of the $D_e$ parameter in family of basis sets obtained by taking out from our actual basis sets two- and one- highest angular momentum functions. Such basis sets roughly correspond to triple- and quadruple-zeta quality (which we denote as TZ and QZ, respectively), while our best basis set is of 5-zeta quality (5Z).

We have found that even for the smallest basis sets of TZ quality the binding energies are very close to the values obtained from those obtained with 5Z basis sets: for the He$^*$+(Li, Na) systems the CCSD(T) interaction energies calculated for the recommended $r_e$'s (Tab.~\ref{MLRvalues}) are, respectively 569.2~cm$^{-1}$ and 350.2~cm$^{-1}$, respectively, that is 6.2$^{-1}$ and 10.6~cm$^{-1}$ below the recommended $D_e$ parameters. Such rapid convergence might result from the fact that in the high-spin He$^*$+alkali-metal systems same-spin electronic pairs are the main contribution to the interaction energy and it is known that the correlation energy for such pairs saturates faster than for the opposite-spin pairs. While we have estimated the CCSD(T) complete basis set (CBS) limit for the He$^*$+(Li, Na) interaction energies to be 576.1~cm$^{-1}$ and 359.3~cm$^{-1}$, respectively, we rather prefer to treat the difference between 5Z basis set and CBS interaction energy as (unsigned) uncertainty attributed to the basis set incompleteness and to take the results for 5Z basis set as recommended values with basis set errors of $\pm$0.6~cm$^{-1}$ and $\pm$1.3~cm$^{-1}$, respectively. 

Our basis set for He$^*$+K is not correlation consistent, so instead of performing the CBS estimate we have simply compared the result in extended ANO-RCC basis set (see Sec. \ref{short-range}) to the interaction energy obtained with basis set with removed $h$ functions. The value of the latter is 468.6~cm$^{-1}$, hence the uncertainty due to basis set incompleteness is about $\pm 1.3$~cm$^{-1}$.

\subsubsection{Post-CCSD(T) contributions to the interaction energy}

Using the CCSDT method in reduced basis sets we have explored the performance of {\em ab initio} methods beyond the CCSD(T) model for He$^*$+(Li, Na) systems using the TZ-quality basis sets. For the systems containing non-polar species quadruply excited configurations give substantially smaller contribution to the binding energies~\cite{smith2014bsc}, hence we might treat CCSDT interaction energy at equilibrium as a probe of post-CCSD(T) effects near the equilibrium distance. Moreover, for the He$^*$+Li case CCSDT is exact if the $1s$ electrons of Li are kept frozen, and is exact also for isolated Li and He$^*$ atoms. We have calculated the CCSDT interaction energy for all electrons active and for the case when $1s$ orbital is frozen. The results are given in the Table \ref{LiNatable}. In a basis set of TZ-quality the CCSDT interaction energy is deeper by merely 3.4~cm$^{-1}$. The TZ basis set is remarkably close to the complete basis set convergence limit (CBS), and we can safely assume that the difference between CCSDT and CCSD(T) is also nearly converged. It is interesting to notice that CCSDT for 3-electron calculations (i.\,e.\, with frozen $1s$ shell of Li) gives actually an even smaller difference between CCSDT and CCSD(T) interaction energies (1.7~cm$^{-1}$), which shows that the core-relaxation effects in this case are comparable to the contributions beyond the CCSD(T) model. 

For He$^*$+Na we were able to calculate the CCSDT interaction energies for 7 active electrons. It turns out that the difference between interaction energies CCSDT and CCSD(T) is only 1.8 cm$^{-1}$ in TZ-quality basis set. We might assume that with converged basis set and all electrons correlated the error might be at most twice as large. 

For He$^*$+K we were not able to converge the CCSDT calculations. Hence, to estimate bounds on $D_e$ we have compared how the analytical long-range potential given by van der Waals series compares to the interaction energies from the actual calculations. It turns out that $C_6$ coefficient extracted from the {\em ab initio} CCSD(T) interaction energies (fitted from 20~$a_0$ to 35~$a_0$) is about 1.5\% smaller compared to the value obtained from perturbation theory (i.\,e.\, Eq.~\ref{C6formula}). If this value is treated as an estimate of {\em ab initio} potential uncertainty it translates to about +7~cm$^{-1}$.

\subsubsection{Bounds on $D_e$ parameters}

As mentioned in Sec.~\ref{short-range}, our methodology predicts that the well depths of the homonuclear dimers He$^*_2$, Li$_2$, Na$_2$, K$_2$ are systematically shallower compared to the experimental values by 0.53\%, 0.92\%, 1.52\% and 2.94\%, respectively. Because of the simple structure of the He$^*$ atom (for isolated He$^*$ atom CCSD(T) method is exact) we expect that within the He$^*$+alkali-metal atom systems the errors should be even smaller. By taking an average of the appropriate percentage uncertainties for homonuclear alkali dimers, i.\,e.\, $\delta_{\rm AB} = (\delta_{\rm A} + \delta_{\rm B} )/2$ we might expect that our {\em ab initio} potentials are deeper by about 0.7\%, 1\% and 1.7\%, which translates to 4.0~cm$^{-1}$, 3.7~cm$^{-1}$, 8.1~cm$^{-1}$ for the He$^*$+(Li, Na, K) potentials, respectively. These uncertainties are consistent with our estimate of post-CCSD(T) interaction energies, which in each case predicts small and systematically positive contributions beyond CCSD(T). We can conservatively assume that the real potentials have a well depth parameter $D_e$ that is: i) not smaller than $D_e$ from CCSD(T) minus basis set uncertainty; ii) not larger than our $D_e$ plus basis set uncertainty, plus the post-CCSD(T) correction. Hence, for He$^*$+(Li, Na, K) systems the $D_e$ parameters for quartet potentials have uncertainties of --1/+4~cm$^{-1}$, --1/+4~cm$^{-1}$ and --1/+9 cm$^{-1}$, respectively.

Finally, we note that nonadiabatic effects can be neglected here despite the relatively small reduced masses. For the He$^*$+Li system in the van der Waals minimum the diagonal Born-Oppenheimer correction calculated with the Hartree-Fock electronic wavefunctions (using the {\sc cfour} program \cite{cfour_brief}) is about 0.02~cm$^{-1}$. For the He$^*$+(Na, K) systems these errors will be even smaller.

\section{Scattering lengths}\label{scatt}

\begin{table}
\caption{Scattering lengths for all He$^*$ + alkali-metal isotopologues, showing the scattering length $a$ corresponding to the recommended $D_e$, and the bounds [$a_-$;$a_+$] corresponding to the bounds on $D_e$. Also the number of bound states $N$ is given. Note that for the alkali-metal atoms the even isotopes are fermions and the odd ones are bosons, while $^4$He is a boson and $^3$He is a fermion.} 
\begin{ruledtabular}
\begin{tabular}{cc|cc|c}
system  		& isotopes  &  $a$		& $[a_-;a_+]$				& $N$    \\  
\hline
He$^*$+Li		& 3 + 6     &  +26 		& $[+23;+26]$ 			&  11      \\  
		       	& 3 + 7     &  $-$17 	& $[-27;-15]$ 			&  11       \\ 
						& 4 + 6     &  +22  	& $[+19;+23]$ 			&  12    \\     
						& 4 + 7     & $-$193 	& $[-607;-161]$  		&  12  \\     
	\hline
He$^*$+Na		& 3 + 23	 	&  +58  	& $[+52;+59]$  			&  11     \\   
						& 4 + 23 	 	&  +7   	& $[-2;+9]$ 				&  12    \\  
	\hline
He$^*$+K		& 3 + 39	  & +51 		& $[+42;+52]$ 			&  13      \\   
						& 3 + 40	  & +49  		& $[+41;+51]$ 			&  13   \\
						&	3 + 41    & +48   	& $[+39;+49]$ 			&  13    \\ 
						&	4 + 39    & +97  		& $[+74;+101]$			&  15     \\
						&	4 + 40    & +91   	& $[+70;+94]$				&  15   \\       
						&	4 + 41    & +86   	& $[+67;+89]$				&  15       \\   
\hline   
He$^*$+Rb \footnote{The scattering length values are shifted by about 1~$a_0$ from our earlier reported values in Ref. \cite{knoop2014umo}, which suffered from a small computational error.}
		& 3 + 85 		&  +5 		& $[-17;+7]$ 				&  13      \\   
						& 3 + 87		&  +3   	& $[-19;+5]$ 				&  13       \\ 
						& 4 + 85  	&  +16  	& $[-4;+18]$ 				&  15    \\     
						& 4 + 87		&  +15  	& $[-6;+17]$ 				&  15  \\  
 \end{tabular}
\end{ruledtabular}
\label{scatteringlengthtable}
\end{table}

With the recommended MLR potentials the quartet scattering length $a$ for all isotopologues can be calculated, for which we have used the 1D renormalized Numerov propagator \cite{johnson1978rnm} for a kinetic energy of 10~nK. Within the Born-Oppenheimer approximation it is only the different reduced masses that give rise to different scattering lengths for the isotopologues within each system and we have verified that non-adiabatic terms are negligible. We can conveniently explore the bounds on the {\em ab initio} scattering lengths solely by scaling the $D_e$ parameter within its estimated bounds. Such scaling does not affect the long-range part of the MLR potential, which is kept fixed. The results are presented in Table~\ref{scatteringlengthtable}, showing $a$ and the bounds $[a_-;a_+]$ related to the bounds on the calculated potentials, where the lowest possible value of $D_e$ corresponds to $a_+$ and the highest one to $a_-$. Also the number of bound states $N$ is indicated, which differs between the two He isotopes by one or two units. 

\begin{figure}
\includegraphics[width=8.5cm]{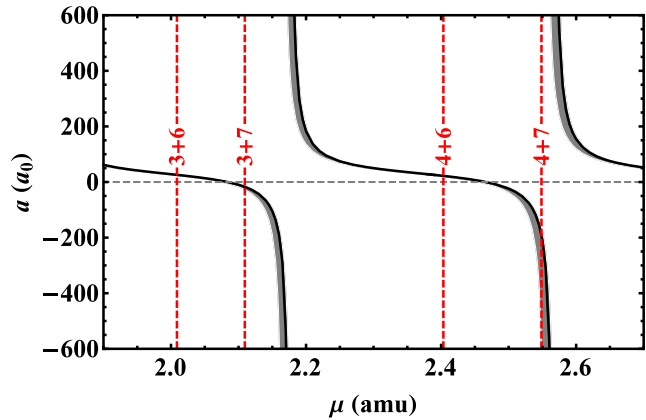}
\caption{(Color online)
\label{massscalingHeLi} Scattering length as function of reduced mass for the He$^*$+Li potential, indicating the reduced masses for the four isotopologues (red dashed lines). The gray area gives the bounds on the scattering length related to the uncertainty in the potential.}
\end{figure} 

The sensitivity of the scattering length $a$ for the potential well depth $D_e$ is very non-linear, and $a$ diverges at each value of $D_e$ at which the potential supports a new bound state. In most cases here $a$ is far away from a pole and the bounds on $a$ are quite tight. A noticeable exception is $^4$He$^*$+$^7$Li, which lies very close to a pole, leading to a broader range of possible scattering length values. This is also illustrated in Fig.~\ref{massscalingHeLi}, showing $a$ as function of the reduced mass $\mu$ for the He$^*$+Li potential. The appearance of poles in $a$ when varying $D_e$ or $\mu$ is similar, as $\mu$ and $V(r)$ appear only as a product in the Schr\"{o}dinger equation (see Eq.~\ref{Schrodinger}). For He$^*$+Li the four isotopologues have quite different reduced masses such that the corresponding scattering lengths can be very different. The similarity between the scattering lengths of $^3$He$^*$+$^6$Li and $^4$He$^*$+$^6$Li is purely accidental. For the heavier alkali-metal atoms (K and Rb) the scattering lengths for the same helium isotope are nearly the same, as the reduced mass hardly changes for the different alkali isotopes. 

The 16 isotopologues of He$^*$+(Li, Na, K, Rb) contain 6 Bose-Bose, 8 Bose-Fermi and 2 Fermi-Fermi mixtures. From the calculated scattering lengths we find that the $^4$He$^*$+($^{23}$Na, $^{39}$K, $^{41}$K, $^{87}$Rb) Bose-Bose mixtures are miscible, while the Bose-Fermi mixtures provide both Fermi core and Fermi shell situations. The $^3$He$^*$+$^{40}$K Fermi-Fermi mixture has already been proposed as its mass ratio is very close to a narrow interval where a purely four-body Efimov effect is predicted \cite{castin2010fbe}. It is interesting to note that both $^3$He$^*$ and $^{40}$K have an inverted hyperfine structure, such that the high-field seeking doubly spin-stretched state combination is the lowest channel within the $^3$He$^*$+$^{40}$K manifold.

From the calculated scattering lengths one can also find whether alkali-metal atoms can be used to sympathetically cool He$^*$, as an alternative to evaporative cooling of $^4$He$^*$ and sympathetic cooling of $^3$He$^*$ by $^4$He$^*$. Although both these schemes are successfully applied, sympathetically cooling with another species would put a less stringent requirements on the initial number of laser-cooled $^4$He$^*$ atoms. Here one has to take into account that the thermalization rate scales with $\xi=4m_{{\rm He}^*}m_{\rm A}/(m_{{\rm He}^*}+m_{\rm A})^2$, and $a^2$ in the zero temperature limit (see for instance Ref.~\cite{knoop2014umo}). Na might be the most suitable candidate for sympathetic cooling of $^3$He$^*$, while $^7$Li is the best coolant for $^4$He$^*$ (although only until quantum degeneracy is reached because of immiscibility). 

\section{Conclusions and outlook}\label{concl}

We have obtained accurate {\em ab initio} \quartet~quartet potentials for He$^*$+(Li, Na, K, Rb), using CCSD(T) calculations and accurate calculations of the $C_6$ coefficients, and have calculated the corresponding scattering lengths for all the isotopologues. An accurate prediction of scattering lengths for these many-electron systems is possible, in contrast to nearly all other types of ultracold mixtures, because of the small reduced masses and shallow potentials that results in a small amount of bound states ($N=11-15$), and therefore a reduced the sensitivity of the scattering length to the properties of the potentials.

So far, we have only considered the quartet potential, which is the only relevant potential for the doubly spin-stretched state combinations, for which Penning ionization is suppressed. Feshbach resonances that allow to tune the scattering length are in principle possible for He$^*$ + alkali-metal systems, as both atoms have electron spin and at least one has nuclear spin, however they would require spin-state combinations in which at least one of the atom is not in the spin-stretched state, and scattering has both doublet and quartet character. Unfortunately, accurate {\em ab initio} calculations of the doublet potentials are far more challenging. First of all, the doublet potentials need multiconfigurational treatment and these methods are at present far less accurate than CCSD(T). Secondly, for the doublet states coupling with continuum states of ionized channels is possible, which might complicate the calculations. In addition, the much larger well depths \cite{ruf1987tio}, and therefore much larger amount of bound states, results in a much stronger sensitivity of the doublet scattering length to the underlying potential. Therefore, experimental input, such as positions of Feshbach resonances, will be needed to determine the doublet scattering properties.

\begin{acknowledgments}
D. K., {\L}. M. and P. S. \.{Z}. acknowledge support from NCN grant DEC-2012/07/B/ST4/01347 and generous amount of CPU time from Wroclaw Centre for Networking and Supercomputing, grant no. 218. S. K. acknowledges financial support by the Netherlands Organisation for Scientific Research (NWO) via a VIDI grant (680-47-511). We thank Mariusz Puchalski for providing the dynamic polarizabilities of metastable helium and Daniel Cocks for corrections on our initial manuscript.

\end{acknowledgments}

\appendix 

\section{Parameters of MLR potential}\label{app}

In Table~\ref{MLRvalues} we give the parameter values for the MLR potentials, where $D_e$ and $r_e$ are obtained directly from the CCSD(T) calculations, $C_6$ is calculated from the dynamical polarizabilities, $C_8$ and $C_{10}$ are taken from Ref.~\cite{zhang2012lri}, and the $\phi$ parameters are obtained from fitting the MLR potential to the CCSD(T) data. The values of $p$ and $q$ are chosen to obtain the best fit.

\begin{table}
\caption{Parameter values of the MLR potentials. The number of digits given exceeds their precision. The $C_8$ and $C_{10}$ coefficients are taken from Ref.~\cite{zhang2012lri}.}
\begin{ruledtabular}
\begin{tabular}{cc|cc}
\multicolumn{4}{c}{\textbf{He$^*$+Li}} \\
\hline
$D_e$  		& 575.40~cm$^{-1}$  													& $\phi_0$ 	& $-$2.5087 \\
$r_e$  	  & 7.5296~$a_0$  															& $\phi_1$  & 0.32009 \\ 
$C_6$ 		& 4.5782$\times 10^{8}$~cm$^{-1}$$a_0^6$ 			& $\phi_2$  & $-$0.38608 \\
$C_8$ 		& 2.9058$\times 10^{10}$~cm$^{-1}$$a_0^8$ 		& $\phi_3$  & 0.34763 \\
$C_{10}$ 	& 2.8093$\times 10^{12}$~cm$^{-1}$$a_0^{10}$	& $\phi_4$  & $-$0.78801 \\
$p$, $q$	&	4, 4																				&	$\phi_5$	& $-$1.3009				\\ 
\hline
\multicolumn{4}{c}{\textbf{He$^*$+Na}} \\
\hline
$D_e$  		& 360.80~cm$^{-1}$  													& $\phi_0$ 	& $-$1.8898 \\
$r_e$  	  & 8.5678~$a_0$  															& $\phi_1$  & 0.30840 \\ 
$C_6$ 		& 4.7723$\times 10^{8}$~cm$^{-1}$$a_0^6$ 			& $\phi_2$  & $-$0.059153 \\
$C_8$ 		& 3.4084$\times 10^{10}$~cm$^{-1}$$a_0^8$ 		& $\phi_3$  & 0.20557 \\
$C_{10}$ 	& 3.4458$\times 10^{12}$~cm$^{-1}$$a_0^{10}$	& $\phi_4$  & $-$0.86720 \\
$p$, $q$	&	5, 5																				&	$\phi_5$	& $-$0.47504				\\
\hline 
\multicolumn{4}{c}{\textbf{He$^*$+K}} \\
\hline
$D_e$  		& 469.83~cm$^{-1}$  													& $\phi_0$ 	& $-$1.8233 \\
$r_e$  	  & 9.08326~$a_0$  															& $\phi_1$  & 0.33689 \\ 
$C_6$ 		& 7.7314$\times 10^{8}$~cm$^{-1}$$a_0^6$ 			& $\phi_2$  & $-$0.12801 \\
$C_8$ 		& 6.7049$\times 10^{10}$~cm$^{-1}$$a_0^8$ 		& $\phi_3$  & 0.13803 \\
$C_{10}$ 	& 7.6487$\times 10^{12}$~cm$^{-1}$$a_0^{10}$	& $\phi_4$  & $-$0.67352 \\
$p$, $q$	&	5, 5																				&	$\phi_5$	& $-$0.22291					\\ 
\hline
\multicolumn{4}{c}{\textbf{He$^*$+Rb}} \\
\hline
$D_e$  		& 452.71~cm$^{-1}$  													& $\phi_0$ 	& $-$1.8284 \\
$r_e$  	  & 9.4079~$a_0$  															& $\phi_1$  & 0.48678 \\ 
$C_6$ 		& 8.4673$\times 10^{8}$~cm$^{-1}$$a_0^6$ 			& $\phi_2$  & $-$0.065081 \\
$C_8$ 		& 8.0108$\times 10^{10}$~cm$^{-1}$$a_0^8$ 		& $\phi_3$  & $-$0.30087 \\
$C_{10}$ 	& 9.4242$\times 10^{12}$~cm$^{-1}$$a_0^{10}$	& $\phi_4$  & $-$1.5195 \\
$p$, $q$	&	5, 4																				&						&					\\ 
\end{tabular}
\end{ruledtabular}
\label{MLRvalues}
\end{table}

\bibliography{HeAlkaliLib}

\end{document}